\documentclass[aps,prl,superscriptaddress,twocolumn]{revtex4-1}
\usepackage{subfigure,graphicx} 
\usepackage{amsfonts,amssymb,amsmath}
\usepackage{flafter}
\usepackage{multirow}
\usepackage{txfonts}
\usepackage[unicode,breaklinks]{hyperref}
\usepackage{siunitx}
\newcommand{\papertitle}{False vacuum decay in an ultracold spin-1 Bose gas}
\hypersetup{
    unicode=true,
    a4paper=true,
    plainpages=false,
    pdftitle={\papertitle},
    pdfauthor={Thomas P. Billam, Kate Brown, Ian G. Moss},
    pdfsubject={\papertitle},
    colorlinks=true,
    linkcolor=blue,
    citecolor=blue,
    filecolor=black,
    urlcolor=blue
}

\usepackage{ulem}

\begin{document} 

\title{\papertitle}

\author{Thomas P.\ Billam}
\email{thomas.billam@ncl.ac.uk}
\affiliation{Joint Quantum Centre (JQC) Durham--Newcastle, School of Mathematics, Statistics and Physics, 
Newcastle University, Newcastle upon Tyne, NE1 7RU, UK}

\author{Kate Brown}
\email{k.brown@ncl.ac.uk}
\affiliation{School of Mathematics, Statistics and Physics, 
Newcastle University, Newcastle upon Tyne, NE1 7RU, UK}

\author{Ian G. Moss}
\email{ian.moss@ncl.ac.uk}
\affiliation{School of Mathematics, Statistics and Physics, 
Newcastle University, Newcastle upon Tyne, NE1 7RU, UK}

\date{\today}

\begin{abstract}{\noindent
We propose an ultracold atom analogue of false vacuum decay using all
three states of a spin-1 Bose gas. We consider a one-dimensional system
with both radio frequency and optical Raman
coupling between internal states. An advantage of our proposal is the lack of a time-modulated
coupling, which can lead to instabilities. Within the elaborate phase structure
of the system we identify an effective Klein-Gordon field and use
Gross-Pitaevskii simulations within the truncated Wigner approximation to model
the decay of a metastable state. We examine the dependence of the rate of vacuum
decay on particle density for $^{7}$Li and $^{41}$K and find reasonable
agreement with instanton methods. 
}
\end{abstract}

\maketitle

First-order phase transitions, characterised by metastable, supercooled states
and the nucleation of bubbles, form an important class of physical phenomena.
In extreme cases, supercooling can lead to a zero-temperature, `false vacuum'
state, and the subsequent decay of the false vacuum via quantum tunnelling
~\cite{Coleman:1977py,Callan:1977pt, Coleman:1980aw}.

The non-perturbative description of vacuum decay involves an {\it instanton},
or \textit{bounce}, solution to the field equations in imaginary time
\cite{Coleman:1977py, Callan:1977pt, Coleman:1980aw}.  \textcolor{black}{
However, the instanton approach gives limited information about how the bubbles emerge in real-time,
and how bubble nucleation events are corellated. 
A recent suggestion has been to explore the details of false vacuum decay} in ultracold atom 
systems, where the impressive
degree of experimental control available raises the
possibility of engineering (possibly
quasi-relativistic) false vacua.  The scheme of Fialko et al.  \cite{FialkoFate2015,FialkoUniverse2017}
is one such proposal. This uses a two-component Bose gas in one dimension
formed from two spin states of a spinor condensate, coupled by a time-modulated
microwave field. After time-averaging, one obtains an effective description
containing a metastable false vacuum state in addition to the true vacuum
ground state.

Refs.~\cite{FialkoFate2015,FialkoUniverse2017,Braden:2017add,Braden:2018tky,HertzbergQuantitative2020}
studied the decay of the false vacuum using field-theoretical instanton
techniques and numerical simulations based on the truncated Wigner
methodology~\cite{Steel1998,blakie_dynamics_2008}.  However,
Refs.~\cite{Braden:2017add,Braden:2019vsw} showed that the false vacuum state
in this scheme can suffer from a parametric instability caused by the
time-modulation of the system. The instability causes decay of the false vacuum
state by a different mechanism than a first-order phase transition.  This
instability presents a challenge to experimental implementation of the scheme
\cite{Braden:2017add,Braden:2019vsw,Billam:2020xna}. Furthermore, the scheme
requires inter-component interactions to be small compared to intra-component
interactions; this necessitates working close to a Feshbach
resonance \cite{FialkoFate2015,FialkoUniverse2017}, which limits flexibility in
the experimental setup.

Clearly, it would be desirable to have an ultracold atom system that simulates vacuum decay
while being free of the need to time-modulate the microwave field and, ideally,
more flexible in terms of experimental setup. In this Letter we show that this
can be achieved using a spin-1 Bose Einstein condensate system with external
coupling fields.  By careful choice of couplings, the system we propose
undergoes vacuum decay in a way that is analoguous to a Klein-Gordon system.

\begin{center}
\begin{figure}[htb]
  \includegraphics[width=0.5\columnwidth]{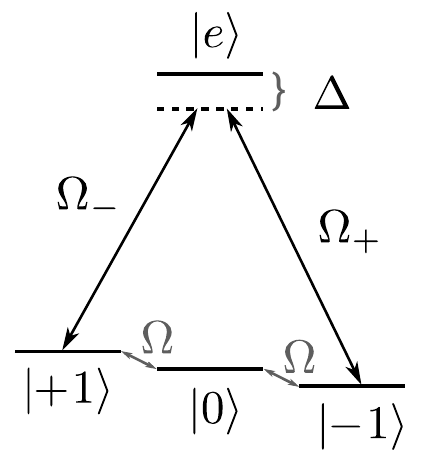}
\caption{Level coupling diagram for the simplest $\Lambda$ system based on
Raman and microwave induced transitions. The $F=1$ spin states labelled
$|m\rangle$, are coupled by a resonant RF beam of frequency $\omega_z$, with
Rabi frequency $\Omega$, and by a two-photon Raman coupling induced by
off-resonant optical beams with Rabi frequencies $\Omega_{\pm}$, zero
two-photon detuning, and detuning $\Delta$ from the excited state $| e
\rangle$.
\label{pot}}
\end{figure} 
\end{center}

We will describe the system in one dimension (1D), assuming the atoms to be
tightly harmonically confined in the transverse directions such that a quasi-1D
description is suitable. The description generalises to two or three
dimensions. We consider a condensate of alkali atoms in their $F=1$ hyperfine
ground state manifold. The degeneracy between internal spin states $|m
\rangle$, where $m\in\{-1,0,1\}$, is lifted by a static magnetic field $B_z$
along the $z$ axis. In addition to intrinsic collisional coupling between the
spin states, described by a quartic Hamiltonian, we propose the states be
extrinsically coupled by both radio frequency fields (RF coupling) and by
optical fields in a two-photon Raman scheme (Raman coupling).

The quadratic terms in our mean field Hamiltonian are
\begin{equation}
H_2=\int dx\left\{
\overline\psi\left[{-\hbar^2\nabla^2\over 2m}-\mu\right]\psi+\overline\psi H^{ZE}_B\psi
+\overline\psi H^{MIX}_B\psi.
\right\},
\end{equation}
where the field $\psi$ has components $\psi_m$.  The constant magnetic field
produces a first order Zeeman effect with frequency $\omega_z=g_F \mu_\mathrm{B} B_z / \hbar$ and a
second order Zeeman effect with frequency $\omega_q$,
\begin{equation}
H^{ZE}_B= \hbar\omega_z J_z+\hbar\omega_q\,J_z^2,
\end{equation}
where $J_x$, $J_y$ and $J_z$ are the dimensionless angular momentum generators.  The RF field has
frequency $\omega_z$ and is polarised in the $x$ direction.  This directly couples
states with azimuthal quantum numbers $m\leftrightarrow m\pm1$.  Coupling of
the $m\leftrightarrow m\pm2$ states can be achieved by two optical fields
arranged on the $D_1$ line, creating a two-photon Raman coupling between the
states in a three-level $\Lambda$ scheme, as shown in Fig.~\ref{pot}. In
presenting our system we neglect complications arising from other states in the
upper hyperfine manifold, and consider only a single excited state $|e \rangle$
with azimuthal quantum number zero. To avoid population of $|e\rangle$, the
detuning $\Delta$ should be large compared to relevant atomic linewidths, and
to keep the momentum transferred to the atoms negligible the optical fields
driving $\sigma_\pm$ transitions should be co-propagating in the $z$-direction \cite{wright_raman}.
We assume zero two-photon detuning. After time-averaging of the RF and optical
frequencies in the rotating wave approximation, we obtain the mixing part of
the Hamiltonian
\begin{equation}
H^{MIX}_B=\frac12\hbar\Omega J_x+\frac12\hbar\alpha \left(J_+^2+J_-^2\right),
\end{equation}
where the frequency $\Omega=g_F \mu_\mathrm{B} B_x / \hbar$ depends on the RF field amplitude $B_x$,
and $\alpha=-\Omega_+\Omega_-/4\Delta$ is determined by the optical-field Rabi
frequencies $\Omega_\pm$ and the detuning $\Delta$.

We assume that the atomic collisions are described by rotationally invariant
dipole-dipole interactions $(\overline\psi\psi)^2$ and $(\overline\psi {\bf
J}\,\psi)^2$,  which we would expect to describe a whole range of systems with
low to moderate external magnetic fields \cite{Kawaguchi2012,Stamper-Kurn2013}.
The interaction terms can be gathered together into an interaction potential
function $V$, so that the total Hamiltonian becomes
\begin{equation}
H=\int dx\left\{
\overline\psi\left[{-\hbar^2\nabla^2\over 2m}\right]\psi+V(\bar\psi,\psi)
\right\}.
\end{equation}
where
\begin{align}
V&=-\mu\overline\psi\psi+\hbar\omega_q(\overline\psi J_z^2\psi)
+\frac12g(\overline\psi\psi)^2+\frac12g'(\overline\psi {\bf J}\,\psi)^2\notag\\
&+\frac12\hbar\Omega\overline\psi J_x\psi+
\frac12\hbar\alpha\overline\psi \left(J_+^2+J_-^2\right)\psi.
\end{align}
Here, $g = 2\hbar \omega_r (a_0 + 2 a_2) /3$ and $g' = 2\hbar \omega_r (a_2 -
a_0) / 3$ where $a_F$ is the $s$-wave scattering length for total-spin-$F$
channels \cite{Kawaguchi2012,Stamper-Kurn2013}, and $\omega_r$ is the trap
frequency of the symmetric transverse confinement. Note that the linear Zeeman
term is cancelled out by the RF field in the rotating wave approximation, and
the magnetisation is not conserved due to mixing. The appropriate
treatment of the spin-1 system for our purposes is one with a fixed chemical
potential but no additional Lagrange multiplier for the magnetisation.

\begin{center}
\begin{figure}[htb]
\begin{center}
 \includegraphics[width=0.8\columnwidth]{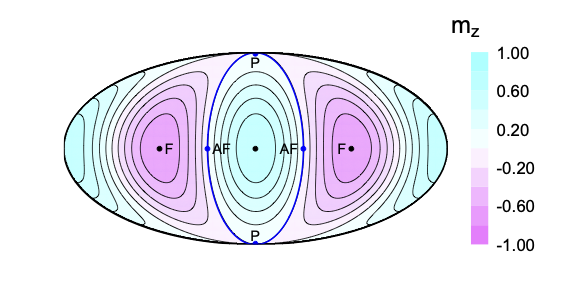}
\end{center}
\caption{Mollweide projection of the sphere $\zeta_0^2+\zeta_{+1}^2+\zeta_{-1}^2=1$ 
($\zeta_0$ in the vertical direction) showing the ground states
superimposed on contours of constant magnetisation. The blue line
represents the locus of BA vacua ($\zeta_+=\zeta_-$) for different external magnetic field strengths.
}
\label{mol}
\end{figure}
\end{center} 

The spin-1 system has a rich phase structure, even in the absence of mixing terms.
Following Kawaguchi and Ueda \cite{Kawaguchi2012}, the fields can be parameterised by
\begin{align}
\psi_{\pm1}&=\sqrt{\rho}\,\zeta_{\pm1}\,e^{i(\theta\pm\varphi)},\label{param}\\
\psi_0&=\sqrt{\rho}\,\zeta_0,
\end{align}
subject to $\zeta_0^2+\zeta_{+1}^2+\zeta_{-1}^2=1$.
The configuration space is the quadrant $\zeta_0>0$, $\zeta_->0$
of  Fig. \ref{mol} and $0<\theta<\pi$, $0<\varphi<\pi$. The ferromagnetic phases (F)
are characterised  by having magnetisation $m_z=\zeta_+^2-\zeta_-^2=\pm 1$.
The other phases have zero magnetisation in our system, and they are
the antiferromagnetic (AF) phase with $\zeta_0=0$, the polar (P) phase with
$\zeta_0=1$ and the broken axisymmetric phase (BA). These phases have
been observed experimentally in ${}^{87}{\rm Rb}$ \cite{Chang_2005}.

\begin{center}
\begin{figure}[htb]
\begin{center}
 \includegraphics[width=\columnwidth]{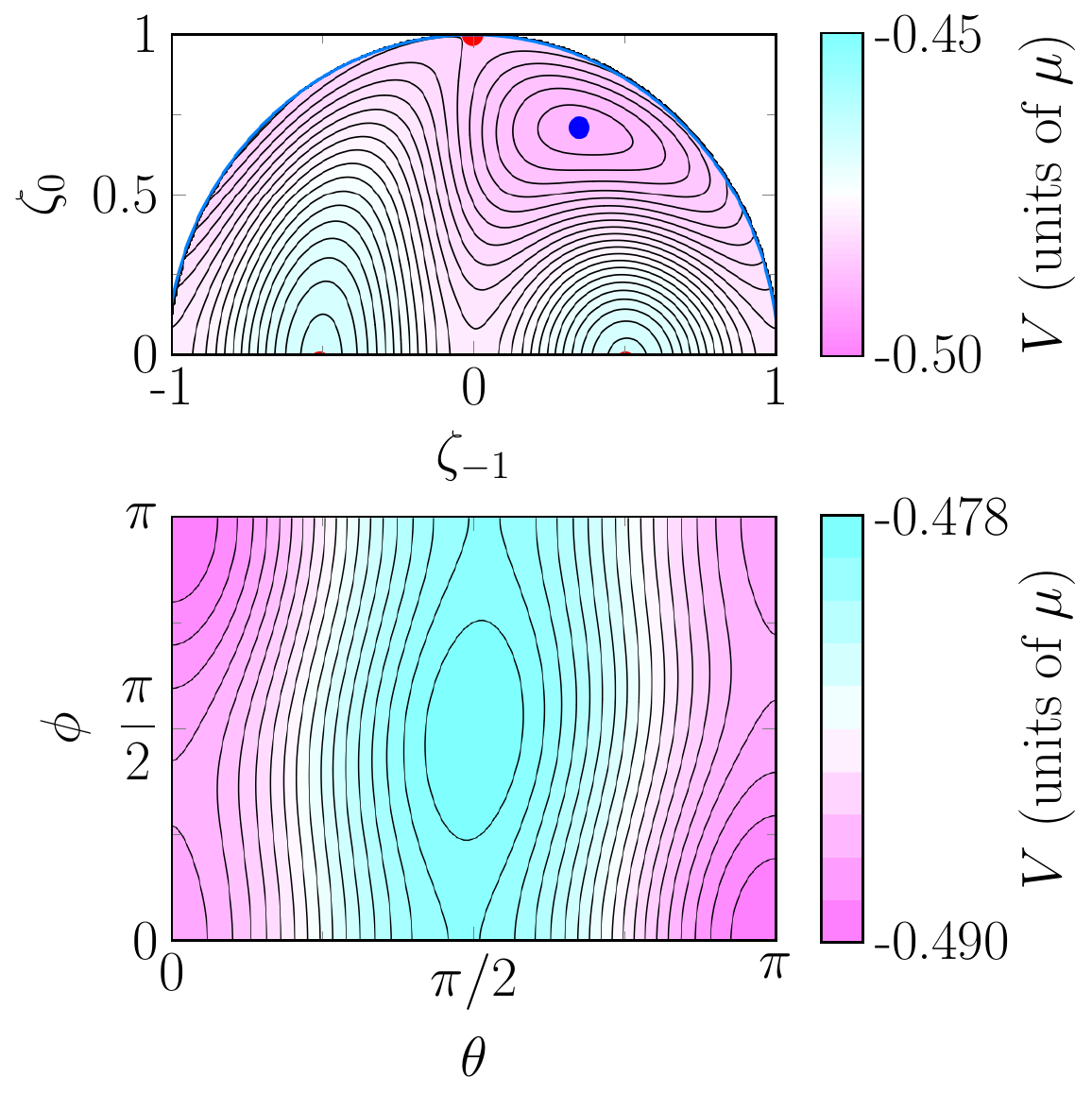}
\end{center}
\caption{The false vacuum, with RF and Raman mixing terms included,
is shown in different projections of the potential $V$ (in units of $\mu$).
The top figure shows the potential as function of the relative amplitudes of the spin components
in a quadrant of the Mollweide projection at $(\theta,\varphi)=(0,0)$ and fixed $\mu$.
The BA vacuum state is indicated by a blue dot.
Below, the potential as a function of the phase angles at $\zeta_\pm=\zeta$.
The false vacuum is at $(\theta,\varphi)=(0,0)$ and the true vacuum at $(\pi,0)$.
In this example, $g'=-0.0256g$, $\epsilon=0.05$, $\lambda=1.7$ and $\omega_q=0.017\mu/\hbar$
(see Table \ref{tab:constants}).
}
\label{vacV}
\end{figure}
\end{center}

We will focus on the BA phase which has the lowest energy when $g'<0$, $g>0$ and 
$0<\hbar\omega_q<-2g'\rho$. If the mixing terms are absent, then
$\zeta_{+1}=\zeta_{-1}=\zeta$ at the minimum, where
\begin{equation}
\zeta=\frac12\left(1+{\hbar\omega_q\over 2g'\rho}\right)^{1/2}.
\end{equation}
Furthermore, we work in the regime $|g'/g| \ll 1$, where $\mu \approx g \rho$,
and also in the regime of weak mixing ($|\hbar\Omega|\ll\mu$) in which the
states have approximately the same moduli as above. Crucially, however, the
weak mixing terms raise the degeneracy between different values of the phase so
that there are stationary points when $(\theta,\varphi)$ equals $(0,0)$,
$(\pi,0)$, $(0,\pi)$ and $(\pi,\pi)$.  The second derivatives of the potential
imply that the stationary points become local minima when
$|\hbar\Omega|\lesssim-2g'\rho$ and $|\Omega|\lesssim-4\alpha$, as shown in
Fig. \ref{vacV}.

Before considering in the dynamics of the phases $\theta$ and $\varphi$ it proves
convenient to rescale the system to natural units. The healing length 
$\xi=\hbar/(mg\rho)^{1/2}$ and natural frequency $\omega_0=g\rho/\hbar$ are defined 
in the usual way. We use the healing length as the length unit, 
$1/\omega_0$ as the time unit and $g\rho$ as the energy unit.
Dimensionless parameters $\epsilon$ and $\lambda$ describe the strength of the mixing terms,
$\epsilon^2=\hbar\Omega/g\rho$ and $\lambda^2=\Omega_+\Omega_-/\Omega\Delta$.

The Klein-Gordon mode can be isolated by fixing $\rho$ and taking 
$\zeta_\pm=\zeta e^{\pm\sigma/2}$. The potential barrier heights in the $\theta$ and $\varphi$ 
directions depend on $g'/g$ and $\epsilon$ respectively.
We take the case where the barrier is smaller in the $\varphi$ directions, i.e. $\epsilon^2\ll |g'/g|$.
The effective Lagrangian density ${\cal L}_{\rm eff}$ at  $O(\epsilon^2)$ then describes a
Klein-Gordon field $\varphi$ with effective Lagrangian,
\begin{equation}
{\cal L}_{\rm eff}=2\zeta^2\rho\left\{
\frac1{2c^2}(\partial_t\varphi)^2-\frac12(\nabla\varphi)^2
-V_{\rm eff}(\varphi)\right\}.
\end{equation}
The propagation speed of the Klein Gordon field is $c$, where $c^2=\omega_q/2$
in healing length units. The potential $V_{\rm eff}(\varphi)$ is
\begin{equation}
V_{\rm eff}=\epsilon^2\lambda_c^2\cos\varphi+\frac12\lambda^2\epsilon^2\sin^2\varphi,
\end{equation}
where
\begin{equation}
\lambda_c=\left({1-g\omega_q/2g'\over
1+g\omega_q/2g'}\right)^{1/2}.
\end{equation}
The potential has a true vacuum at $\varphi=\pi$ and a false vacuum at $\varphi=0$ provided that
$\lambda>\lambda_c$. The effective Klein-Gordon field has a mass 
$m_\varphi=\epsilon(\lambda^2-\lambda_c^2)^{1/2}$ in the false vacuum.

\begin{center}
\begin{figure}[htb]
\begin{center}
 \includegraphics[width=\columnwidth]{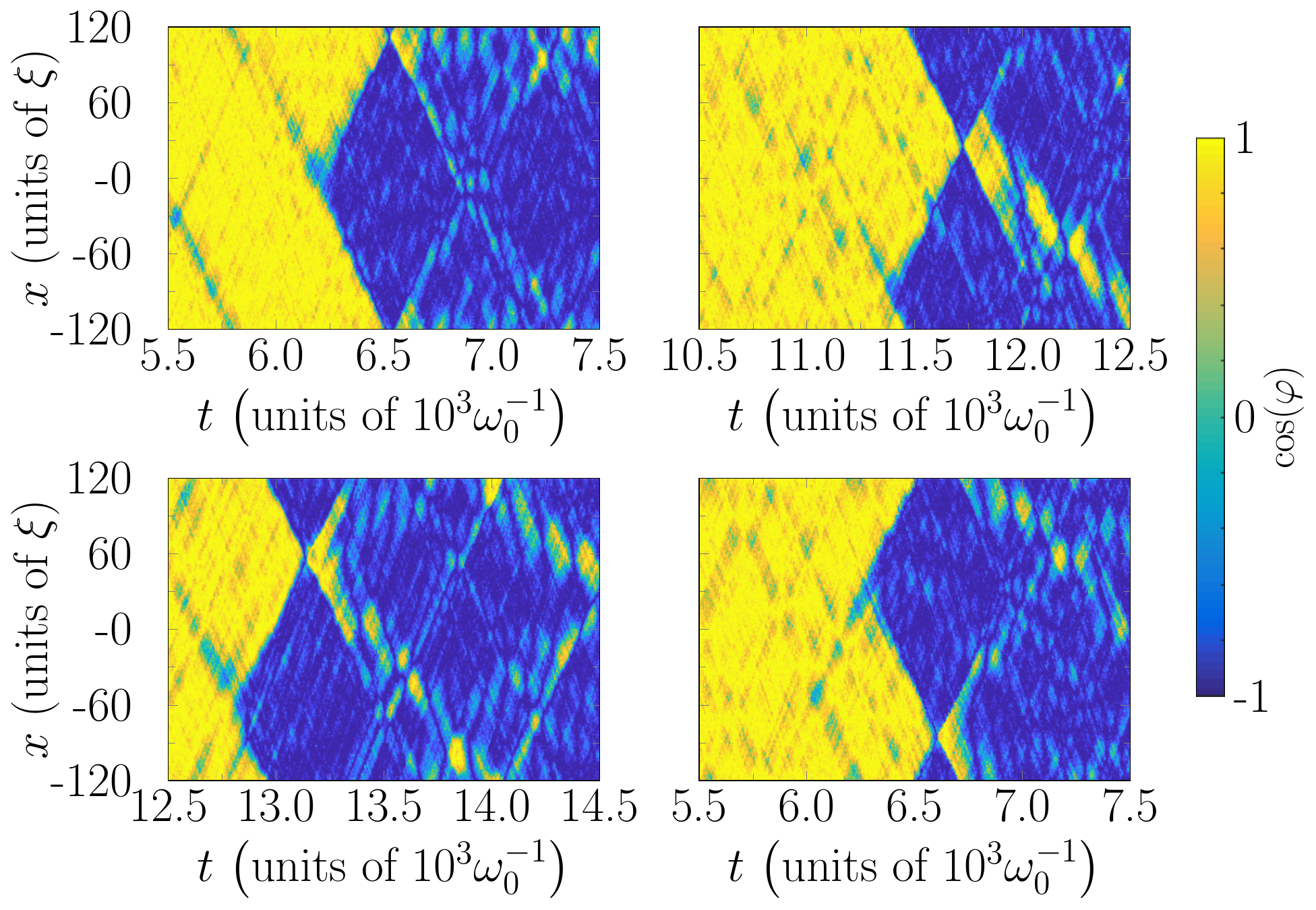}
\end{center}
\caption{Example trajectories showing bubble nucleation in $^7$Li (see
Table~\ref{tab:constants}) with dimensionless coupling parameters $\lambda =
1.7$ and $\epsilon = 0.05$, and dimensionless density $\rho \xi = 20$.
}
\label{images}
\end{figure}
\end{center}

We now perform numerical simulations using the projected Gross-Pitaevskii equation (PGPE) in the 
truncated Wigner approximation (TWA) to examine how the decay of the false vacuum proceeds in the 
fully non-linear system. \textcolor{black}{On the face of things, the existing
theory behind the TWA approach does not extend to non-perturbative quantum
phenomena. However, numerical simulations carried out on related systems have
shown remarkable agreement between the vacuum decay obtained from TWA and
bubble nucleation obtained from the instanton approach
\cite{FialkoFate2015,FialkoUniverse2017,Braden:2017add,Braden:2018tky,HertzbergQuantitative2020}.
We therefore proceed to compare the two approaches in the spin-1 system.
}

We take a one dimensional system with
periodic boundary conditions, such as would be seen in an ultracold atom ring trap.
Including the projector, the dimensionless PGPE reads 
\begin{equation}
  i \, {\partial \psi\over\partial t}
  = \mathcal{P} \left\{
-\frac12{\partial^2\psi\over \partial x^2}+{\partial V\over\partial \overline \psi} \right\},
  \label{spgpe_with_p}
\end{equation}
In order to represent the quantum fluctuations, we take a stochastic field $\psi$
initially in the false vacuum state with small fluctuations correlated
to match the linearised quantum system.
Most of the fluctuations are in the Bogliubov modes corresponding to the phase direction, 
and we have placed the noise in these modes only. The relevant sector of 
Bogliubov-de Gennes modes has dispersion relation
\begin{equation}
\omega(k)=\frac12\left(k^2+2\omega_q\right)^{1/2}\left(k^2+m_\varphi^2\right)^{1/2},
\end{equation}
and the fluctuations associated with these modes has power spectrum
\begin{equation}
\langle\varphi_k\varphi_{k'}\rangle=
{1\over 8\rho\zeta^2}\left({k^2+2\omega_q\over k^2+m_\varphi^2}\right)^{1/2}\delta_{kk'}
\end{equation}
This is identical to a Klein-Gordon result in the range $k\ll (2\omega_q)^{1/2}$.
When combined with limits on the quadratic Zeeman shift in the BA vacuum,
fluctuations will appear relativistic when $k\ll 2|g'/g|^{1/2}$.
It follows that it is more difficult to replicate relativistic behaviour in systems with very small
values of $|g'/g|$.

The projection ${\cal P}$ in the PGPE cuts off modes with wave number $k>k_c/2$, where $k_c$
is the largest wave number that can be accommodated on the finite sized grid,
ensuring we can compute the time-evolution of the field using a Fourier
pseudospectral method without any aliasing of the nonlinear terms. In
our simulations we use a $601$-point grid of length $120\,\xi$, and evolve the
equations with a 4th order Runge--Kutta timestep of $dt=10^{-4}\,\omega_0^{-1}$ using XMDS2
software~\cite{DennisXMDS2013}. To reduce the possible parameter space 
we fix the quadratic Zeeman shift to $\omega_q = -2 g' / 3g$ in all
simulations. 

\begin{center}
\begin{figure}[htb]
\begin{center}
 \includegraphics[width=\columnwidth]{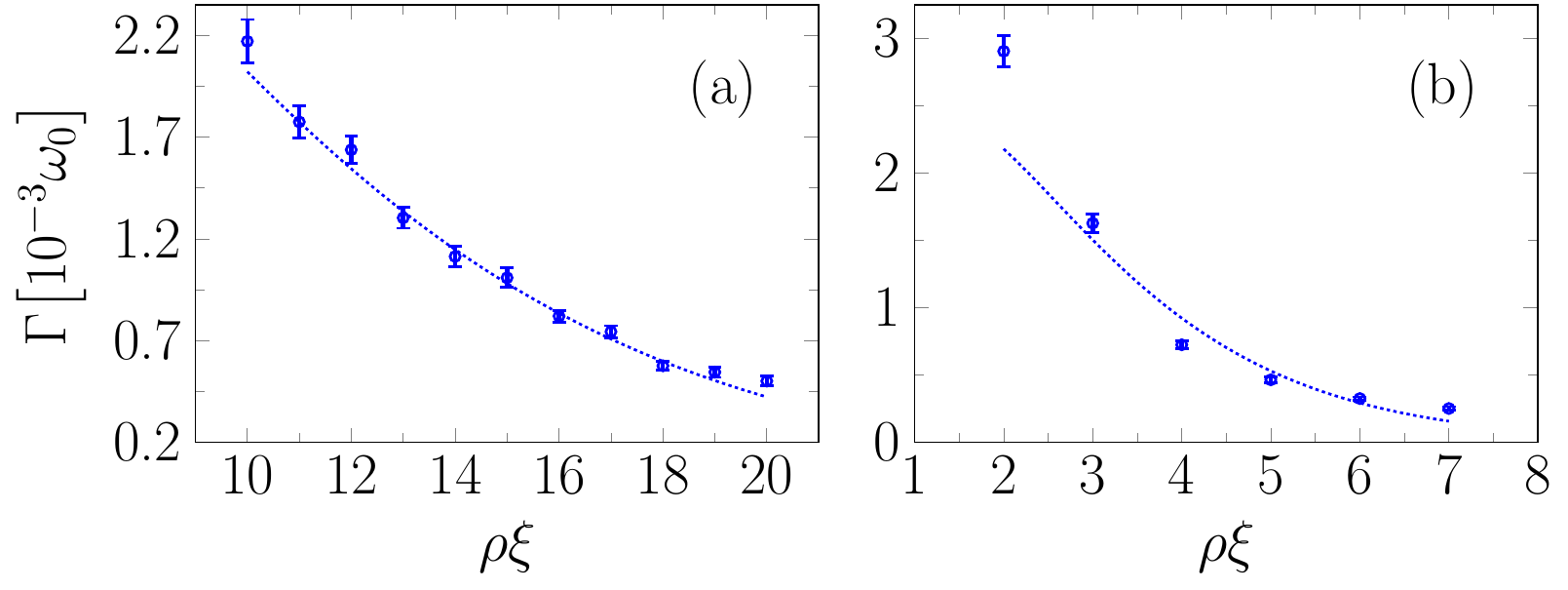}
\end{center}
\caption{The vacuum decay rate $\Gamma$ for (a) $^{7}$Li and (b) $^{41}$K, with
$\lambda=1.7$ and $\epsilon=0.05$, plotted as a function of the particle
density $\rho$. The solid curve represents the decay rate calculated using the
instanton method, after fitting to the prefactor $A$ and an effective coupling
$\lambda_{\rm eff}$. There is good agreement to the relativistic theory for
$^{7}$Li, but less good agreement for $^{41}$K (which has a smaller value of $|g'/g|$).
}
\label{decayrate}
\end{figure}
\end{center}

Our simulations show that the system nucleates false vacuum regions, as in the
examples shown in Fig. \ref{images}. In this periodic simulation, single
bubbles undergo self-collision. In the collision region, there is a sudden
release of energy that locally restores the metastable state for a short while.
The decay is measured by marking the time at which the spatial average
$\langle\cos\varphi\rangle$ becomes larger than $-1+\delta$, where $\delta=0.9$
is chosen to be much larger than the typical fluctuations of
$\langle\cos\varphi\rangle$ due to quantum fluctuations in the system.  Running
many stochastic trajectories allows us to compute the probability, $P$, of
remaining in the metastable state at time $t$. A fit to the exponential form
$P=a e^{-\Gamma t}$ over the time intervals seen to be exhibiting exponential
decay (we find this to be times late enough that $P < 0.7$) yields the decay
rate $\Gamma$. The decay rates for two systems are plotted in Fig. \ref{decayrate},
with error bars estimated using a bootstrap procedure as
described in~\cite{BillamSimulating2019}.  The decay rates have also been compared
to the prediction using Coleman's instanton method,
$\Gamma=ABe^{-B}$, using values for the exponent $B\equiv B(\rho,\lambda)$
taken from \cite{Mario}. In common with our previous work \cite{Billam:2020xna}, 
we found it necessary to modify the effective value of the coupling and replace it with a renormalised
value $\lambda_{\rm eff}$. Treating the prefactor $A$ and the coupling $\lambda$
as free parameters in the fit is an interim measure that could be improved if the radiative corrections to the
tunnelling exponent were known.

\begin{table}
\caption{\label{tab:constants}Physical properties used to compute simulation parameters. Scattering lengths are from the table in Ref.~\cite{Stamper-Kurn2013}.}
\begin{ruledtabular}
\begin{tabular}{lcccl}
Species & $a_0\,(a_\mathrm{Bohr})$ & $a_2\,(a_\mathrm{Bohr})$ & $g'/g$ & $\Delta E_\mathrm{hfs}\,$(MHz) \\
\hline  \\ [-1.5ex]
$^{7}$Li & $23.9$ & $6.9$ & $-0.456$ & $803.5 \times h$ \cite{Li_hyperfine_ref} \\
$^{41}$K & $68.5$ & $63.5$ & $-0.0256$ & $254.0 \times h$ \cite{K_hyperfine_ref_1, K_hyperfine_ref_2}  \\
$^{87}$Rb & $101.8$ & $100.4$ & $-0.0046$ & $6834.7 \times h$ \cite{Rb_hyperfine_ref}
\end{tabular}
\end{ruledtabular}
\end{table}

Finally, we comment on the experimental viability of our system. We tabulate
relevant physical properties for alkali species with the required property $g'/g
< 0$ in Table~\ref{tab:constants}. The ground state hyperfine energy splitting
$\Delta E_\mathrm{hfs}$ determines the magnetic field needed to achieve a given
quadratic Zeeman shift~\cite{Stamper-Kurn2013}. While $g'/g$ is fixed by the
atomic species, there is considerable flexibility in choosing tunable
experimental parameters that correspond to the dimensionless parameters used in
our simulations. As examples, the parameters used in Figs.  \ref{images} and
\ref{decayrate}~(a) with $\rho \xi = 20$ would correspond to $2400$ $^7$Li
atoms in a $\SI{260}{\micro m}$ circumference ring trap with transverse
frequency $\omega_r = 2\pi \times \SI{25}{\kilo Hz}$ and a bias field of
$B_z=0.39\,$Gauss.  The timescale $\omega_0^{-1}$ corresponds to
$\SI{0.52}{\milli s}$.
The parameters used in Fig.~\ref{decayrate}~(b) with
$\rho \xi = 7$ would correspond to $840$ $^{41}$K atoms in a $\SI{24.4}{\micro
m}$ circumference ring trap with the same transverse frequency and a bias field
of $B_z=0.23\,$Gauss. We assume there is very wide experimental flexibility in
terms of the coupling field Rabi frequencies and detuning ($\Omega$,
$\Omega_\pm$, $\Delta$); in practice these would need to be tuned to give the
desired $\epsilon$ and $\lambda$ by taking into account the additional,
smaller, light shifts arising from the other states in the upper hyperfine
manifold that we neglect here.
We note that the example parameter values suggested above would
appear to require a very low temperature ($\SI{2.5}{\nano K}$ for $^{7}$Li and
$\SI{16.6}{\nano K}$ for $^{41}$K) to achieve complete phase coherence across
the system in a single-component Bose gas. Crucially, however, the false vacuum
state exists in the \textit{relative} phase, and the relevant condition is that $T<m_\varphi$
in dimensionless units. This sets the temperatures for
\textit{relative} phase coherence to $\SI{53}{\nano K}$ for $^{7}$Li and
$\SI{352}{\nano K}$ for $^{41}$K. In principle, false vacuum  decay
should be observable in  $^{87}$Rb, but we were unable to find a favourable parameter
regime given the very small $|g'/g|$ ratio.

In conclusion, we identified a metastable state of an effective Klein-Gordon
field in a radio frequency and optical Raman coupled spin-1 Bose gas, which
could serve as a laboratory example of false vacuum decay.  Compared to
previous proposals using same-species two-component Bose gases our
proposal does not require time-modulation of the coupling, thus avoiding
problematic instabilities, and avoids the need to minimize inter-component
scattering length using Feshbach resonances. We numerically characterized false
vacuum decay in the system, finding reasonable agreement with instanton
predictions. Our proposal may provide a practical alternative system in which to realize
an analogue to relativistic false vacuum decay in $^7$Li or $^{41}$K
experiments.

Data supporting this publication are openly available under
a Creative Commons CC-BY-4.0 License in Ref.~\cite{data_package}.

{\em Acknowledgements:}
We would like to thank Jonathan Braden for helpful discussions.
This work was supported by the UK Quantum Technologies for Fundamental Physics 
programme [grant ST/T00584X/1].  KB is supported by an STFC studentship. 
This research made use of the Rocket High Performance 
Computing service at Newcastle University. 

\bibliography{paper}

\end{document}